\documentclass{ar-1col}

\usepackage[comma]{natbib}
\usepackage{url}
\usepackage[font=normal]{caption}
\usepackage{tikz}
\usetikzlibrary{arrows,automata,er,positioning,calc,arrows}
\usetikzlibrary{shapes,decorations,arrows,calc,arrows.meta,fit,positioning}
\tikzset{
	-Latex,auto,node distance =1 cm and 1 cm,semithick,
	state/.style ={ellipse, draw, minimum width = 0.7 cm},
	point/.style = {circle, draw, inner sep=0.04cm,fill,node contents={}},
	bidirected/.style={Latex-Latex,dashed},
	el/.style = {inner sep=2pt, align=left, sloped}
}

\usepackage{amsmath}

\jname{Xxxx. Xxx. Xxx. Xxx.}
\jvol{AA}
\jyear{YYYY}
\doi{10.1146/((please add article doi))}

\begin{document}

\markboth{Gorfine and Zucker}{Shared Frailty Methods in Complex Survival Data}

\title{Shared Frailty Methods for Complex Survival Data: A Review of Recent Advances}

\author{Malka Gorfine$^1$ and David M. Zucker$^2$ 
\affil{$^1$Department of Statistics and Operations Research, Tel Aviv University, Tel Aviv, Israel 6997801; email: gorfinem@tauex.tau.ac.il}
\affil{$^2$Department of Statistics and Data Science, Hebrew University of Jerusalem, Jerusalem, Israel}
}

\begin{abstract}
Dependent survival data arise in many contexts. One context is clustered survival data, where survival data are collected on clusters such as families or medical centers. Dependent survival data also arise when multiple survival times are recorded for each individual. Frailty models is one common approach to handle such data. In frailty models, the dependence is expressed in terms of a random effect, called the frailty. Frailty models have been used with both Cox proportional hazards model and the accelerated failure time model. This paper reviews recent developments in the area of frailty models in a variety of settings. In each setting we provide a detailed model description, assumptions, available estimation methods, and \texttt{R} packages.
\end{abstract}

\maketitle

\tableofcontents

\section{INTRODUCTION}

Clustered time-to-event outcomes arise frequently in various research areas such as epidemiology and economics, among others. The clusters typically are families, clinical centers, schools, or companies. Consider, for example, family studies, where the time to a certain disease diagnosis is recorded for several members in each family. Unobserved genetic and environmental background factors that family members share often leads to correlation among the outcomes within family members, even after accounting for
the observed genetic and environmental information. Introducing random variables to characterize unobserved dependence in the data is a popular practice in regression models. In the context of survival analysis (i.e., time-to-event outcome analysis), such models are known as shared frailty models: the cluster-specific random effect representing the unobserved common risk shared by cluster members is called the frailty. In these models, it is assumed that the random frailty factor captures all the within-cluster
dependence, so that the failure times of cluster members are assumed independent given the observed covariates and the unobserved frailty.

Competing events data arise when individuals are susceptible to several types of events but can experience at most one event  \citep{klein2003survival,kalbfleisch2011statistical}. For example, competing risks for length of hospital stay are discharge and in-hospital death. Occurrence of one of these events precludes us from observing the other event on this patient. Another classical example of competing risks is cause-specific mortality, such as death from heart disease, death from cancer, and death from other causes. Often, one event type is singled out as the event of interest, while the others may prevent the event of interest from occurring. For example, leukemia relapse or AIDS may be unobservable because the patient died before the diagnosis of these events. It is well known that caution is needed in estimating the distribution of the time to the event of interest in the presence of the competing risks.

Related to the competing risks setting the illness-death model setting, also known as the semi-competing risks setting.
This setting involves a nonterminal event and a terminal event, with the terminal event censoring the nonterminal event
but the nonterminal event not censoring the terminal event. In this setting, there is partial information on
the within-individual dependence between the nonterminal event and the terminal event, and frailty models have been used to
model this dependence. This is an example of a situation where the individual is the cluster. Another setting of this type
is the recurrent events setting, where an individual can experience the event of interest multiple times over 
the course of follow-up.

For clustered survival data with a single event type such as death (i.e. no competing risks), there is a rich literature,
including tutorials, with most papers focusing on the Cox-type proportional hazards (PH) model, see \cite{balan2020tutorial} and references therein. The goal of the current review is to cover recent advances in shared-frailty models beyond Cox-type models with a single event type. This includes: (1) Accelerated failure time (AFT) models for clustered data, an important alternative to Cox PH models; (2) The setting of competing events for clustered survival data; (3) Illness-death models for non-clustered data; (4) Recurrent events data; and
(5) Random survival forests applied to clustered data.

This review does not cover two other popular methods for clustered data, standard marginal Cox or AFT regression methods and copula models. Marginal modeling methods typically use a working assumption of independence of outcomes within a cluster;
the within-cluster dependence is accounted for in computing the standard errors for the regression coefficients,
but the within-cluster dependence structure is left unspecified and is not estimated. In copula models, the joint survival function of the cluster members' survival times is modeled as a function, called the copula, of the marginal survival functions of the cluster members' survival times. The copula function is used to couple the marginal survival functions and the joint survival function, and determines the type of dependence. The frailty model is a conditional hazard model, and the joint survival function can be obtained by integrating out the frailty. \cite{goethals2008frailty} explained in depth the similarities and the differences between frailty and copula models.  

The structure of the paper is as follows. Section 2 presents basic notation. Section 3 discusses frailty models for clustered single-event
survival data. Both Cox-type PH models and AFT models are discussed. Section 4 discusses Cox-type frailty models for clustered competing risk
data. Sections 5 and 6 discuss, respectively, frailty models for the illness-death setting and the recurrent events setting. Sectiom 7
discusses the use of frailty models with the random forest approach applied to clustered survival data. Section 8 provides concluding remarks.
In each setting we provide a detailed model description, assumptions, available estimation methods, and \texttt{R} packages.

\section{NOTATION}

Consider $n$ independent clusters, with cluster $i$, $i=1,\ldots,n$, having $m_i \geq 1$ members. For member $k$ of cluster $i$, let $T^o_{ik}$ and $C_{ik}$ be the time to failure and the censoring time, respectively, and let $Z_{ik}$ be a vector of covariates. For simplicity of presentation, we consider only time-independent covariates; extension to the case of time-dependent covariates is straightforward for some models and more complicated for others.
Let $J_{ik} \in \{1, \ldots, L\}$ be the type of failure observed, and let $J_{ik} = 0$ in the case of censoring.
Define $T_{ik} = \min(T^o_{ik}, C_{ik})$ and $\delta_{ik} = I(T^o_{ik} \leq C_{ik})$, and let $\tau$ be the
maximal follow-up time. The frailty may be time-dependent and may be vector-valued.
We express the frailty
process of cluster $i$ as $\omega_i(t) = \exp\{ \epsilon_{i}(t) \}$,
where $\epsilon_i(t) = \{\epsilon_{i1}(t), \ldots, \epsilon_{iL}(t)\}$ is a random process over $[0,\tau]$,
possibly vector-valued. Define $\mathcal{F}_{it}$ to be the $\sigma$-algebra generated by $\epsilon_i(t)$ through its history up to and including time $t$, $\mathcal{F}_{it} = \{\epsilon_i(u), 0 \leq u \leq t\}$. This is a general notation covering the following settings included in this review:
\begin{itemize}
	\item \textit{Clustered data with no competing events} such that $J_{ik} \in \{0,1\}$ and time-independent, scalar-valued frailty. An example is a family study where ages at death are recorded for multiple members of the same family.
	\item \textit{Clustered data with competing events.} One example is a breast cancer family study. The BRCA1 mutation
is known to markedly increase the risk of multiple cancers such as breast and ovarian cancers \citep{risch2006population}. Thus, in estimation of breast cancer risk, a subject may be censored due to death from ovarian cancer that occurred before the onset of breast cancer that might otherwise have occurred. 
	\item  \textit{The semi-competing risks setting via illness-death models.}  This setting involves three stochastic processes (see Figure~1): the time to a pre-specified nonterminal event (e.g. age at disease diagnosis), time to a pre-specified terminal event free of the nonterminal event (e.g. age at death free of the disease), and time to the terminal event since the nonterminal event (e.g. age at death after disease diagnosis). In this setting, clusters are comprised by the individuals
	and the dependent event times $T^o_{i1}$ and $T^o_{i2}$ correspond, respectively, to nonterminal event and terminal events.
If an individual experiences the terminal event before experiencing the nonterminal event, we define $T^o_{i1}=\infty$. Also, the frailty process $\epsilon_i(t)$ is scalar-valued, and the same censoring time $C_i \equiv C_{i1}=C_{i2}$ applies to both events. Event indicators for this setting will be defined in Section 5.
\item \textit{Recurrent events data}: Examples include repeated epileptic seizures and repeated hospitalizations.
\end{itemize}
In all the regression models to be discussed, $\beta$ and $\lambda$ will be used to denote a vector of regression coefficients and a hazard function, respectively, although their practical interpretation varies from one model to the other. Also, $\epsilon_i$ or $\omega_i$ will be used interchangeably to refer to the frailty variate. 

\section{CLUSTERED DATA WITHOUT COMPETING EVENTS}

This section reviews existing shared frailty models and estimation procedures for clustered data with a single event type.
Here, the observed data consists of $\{T_{ik},\delta_{ik},Z_{ik} \, ; \, i=1,\ldots,n \, , \, k=1,\ldots,m_i \}$. 
The recent tutorial of \cite{balan2020tutorial} provides a comprehensive discussion of Cox-type shared frailty models. For
completeness, we provide a brief review of the Cox-type models. We then proceed to a detailed discussion of AFT-type shared frailty
models. 

\subsection{Cox Proportional Hazards Models}
Consider $n$ independent clusters and a time-independent frailty variates, $\omega_i \equiv \omega_{i}(t)$, $i=1,\ldots,n$. The conditional
hazard function for individual $ik$, conditional on the cluster frailty $\omega_i$, is assumed to take the form
$$
\lambda_{ik}(t|Z_{ik},\omega_i) = \omega_i \lambda_0(t) \exp(\beta^T Z_{ik})   \hspace{0.4cm} i=1,\ldots,n \hspace{0.4cm} k=1,\ldots,m_i
$$
where $\lambda_0$ and $\beta$ are, respectively, an unspecified baseline hazard function and a $p$-vector of unknown regression coefficients.
This is a popular extension of the \cite{cox1972regression} proportional hazards model where a common random effect acts multiplicatively on the hazard rates of
all cluster members. Individuals in clusters with a large value of the (unobserved) frailty $\omega_i$ will 
tend to experience the event at earlier times than clusters with a small value. 

\textit{Assumptions:} In addition to the standard regularity assumptions often used with Cox proportional hazards model \citep{fleming1991counting}, the frailty-specific assumptions usually are: (A.1) The frailty $\omega_i$ is independent 
of $Z_{ik}$ and follows a cumulative distribution function $F$ indexed by an
unknown parameter $\theta$. (A.2) The frailty distribution $F$ is chosen such that the hazard model is identifiable;
this can be done, for example, by using a distribution with expectation fixed at 1. (A.3) Given $Z_{ik}$ and $\omega_{i}$, the censoring is independent and noninformative for $\omega_i$ and $(\beta, \lambda_0)$, in the sense of \citet[Section III.2.3]{andersen1993statistical}. 

\textit{Estimation:} Early publications tended to focus on maximum likelihood (ML) estimators and the EM algorithm under
a specified frailty distribution. A partial list of distributions considered includes the one-parameter gamma \citep{clayton1978model,gill1985discussion,nielsen1992counting,klein1992semiparametric,parner1998asymptotic}, the positive stable \citep{hougaard1986class}, the inverse Gaussian \citep{hougaard1986survival}, the lognormal \citep{mcgilchrist1991regression}, and the power variance family \citep{aalen1988heterogeneity,aalen1992modelling}. \cite{therneau2003penalized} showed that ML
estimation for the gamma frailty model can be cast in a penalized likelihood framework.
\cite{ripatti2000estimation} presented an approximate ML procedure of penalized likelihood form for the lognormal frailty
model. \cite{ha2001hierarchical,do2017statistical} followed the hierarchical-likelihood (h-likelihood) approach
of \cite{lee1996jrssb}, which
treats the latent frailties as parameters and estimates them jointly with other parameters of interest. \cite{jeon2012bias} showed that the h-likelihood estimators perform well when the censoring rate is low but are substantially biased when the censoring rate is moderate to high, and proposed a bias-correction method for the h-likelihood estimators under a shared frailty model. \cite{jeon2012bias} also extended the h-likelihood approach to accommodate a vector of frailty variates, which covers, 
for example, the case where each cluster is further divided into correlated subgroups.

Later, \cite{gorfine2006prospective}  provided a pseudo-likelihood procedure that handles any parametric frailty distribution with finite moments (that satisfies identifiability conditions). \cite{zeng2007maximum} presented an elegant and effective method to calculate the semiparametric maximum likelihood estimators under any parametric frailty distribution (that satisfies identifiability conditions) and under a new class of transformation models.  They recommended the lognormal frailty over gamma. Both the estimation procedure of \cite{gorfine2006prospective} and that of \cite{zeng2007maximum} were shown to provide consistent and asymptotic normal estimators. 

\textit{Software:} Available R packages \citep{team2013r} are listed in Table 1. Earlier works and packages use the gamma frailty distribution because of its mathematical tractability. The gamma model, however, induces a restrictive form of dependence. We encourage practitioners to take advantage of other available frailty distributions. While diagnostics for frailty models have received limited attention  
\citep{shih1995assessing,glidden1999checking,cui2004checking,munda2013diagnostic,geerdens2013goodness}, \cite{glidden2004modelling,hsu2007effect,gorfine2012conditional}  demonstrated by extensive simulation study that the gamma frailty model is robust to frailty distribution mis-specification 
with respect to both bias and efficiency loss in the estimation of the regression parameters.  


\subsection{AFT Models}
Although the \cite{cox1972regression} proportional hazards models are highly popular, AFT models offer two main advantages: (i)  The log-linear
formulation provides a simple interpretation of the regression parameters \citep{keiding1997role}. (ii) The estimates of the regression parameters are robust to omitted covariates \citep{hougaard1999fundamentals}, in contrast to  proportional hazards models \citep{hougaard1994heterogeneity}. A standard AFT model that ignores possible dependence between cluster members' failure times is given by
$$
\log T^o_{ik} = \beta^T Z_{ik} + \sigma U_{ik}  \hspace{0.4cm} i=1,\ldots,n \hspace{0.4cm} k=1,\ldots,m_i
$$
where the subject-specific error terms $U_{ik}$ are independent and identically distributed. The most common parametric
AFT models take the distribution of $U_{ik}$ to be fully specified (e.g., standard normal or extreme value) without
unknown parameters, so that the only unknown parameter aside from $\beta$ is the dispersion parameter $\sigma$. 
It is also possible to assume that the distribution of $U_{ik}$ follows a parametric distribution with
unknown parameters to be estimated. A nonparametric AFT approach leaves the distribution of $U_{ik}$ unspecified (and 
then $\sigma$ can be omitted). \cite{lambert2004parametric} considered the shared frailty AFT model 
$$
\log T^o_{ik} = \beta^T Z_{ik} + \epsilon_i + \sigma U_{ik}  \hspace{0.4cm} i=1,\ldots,n \hspace{0.4cm} k=1,\ldots,m_i
$$
Under this model,
\begin{equation}
	\lambda_{ik}(t|Z_{ik},\epsilon_i) = \lambda_0(t\exp\{-\beta^T Z_{ik} - \epsilon_i \})  \exp(-\beta^T Z_{ik} - \epsilon_i)
\label{eq:additive}
\end{equation}
where $\lambda_0$ is the hazard function of $\exp(\sigma U_{ik})$. Here, $\epsilon_i$ mirrors the role played by the 
frailty term in the extended Cox proportional hazards model discussed in the preceding subsection.
However, under the AFT model the term $\epsilon_i$ is often called a ``fortitude" term,
since, in contrast with the extended Cox models, larger values of $\epsilon_i$ correspond to increased lifetimes.

Under Assumptions A.1--A.3 above, \cite{lambert2004parametric} considered maximum likelihood estimators (by integrating out the
unobserved frailties)
under parametric choices for the baseline hazard function $\lambda_0$ and the distribution of the frailty component. The estimation
procedure can be carried out using optimization and numerical integration routines.

\cite{pan2001using} adopted a different modeling approach, in which the frailty is incorporated into the error term of the AFT model. Specifically, it is assumed that
$$
\log T^o_{ik} = \beta^T Z_{ik} +  U_{ik}  \hspace{0.4cm} i=1,\ldots,n \hspace{0.4cm} k=1,\ldots,m_i
$$
where the conditional hazard function of $U_{ik}$ (on the entire real line)
given the cluster-level frailty variate $\omega_i$ is given by
$$
\lambda_{ik}(t|\omega_i) = \omega_i \lambda_0(t)
$$
and $\lambda_0$ is an unspecified baseline hazard function (and thus $\sigma$ is omitted). Under this model, the conditional hazard functions, given $(Z_{ik},\omega_i)$, becomes
\begin{equation}\label{eq:pan2001}
\lambda_{ik}(t|Z_{ik},\omega_i) = \omega_i \lambda_0(t \exp\{- \beta^T Z_{ik} \}) \exp(-\beta^T Z_{ik}) \, ,
\end{equation}
so that Equation \ref{eq:pan2001} involves a the multiplicative frailty effect, in contrast to Equation \ref{eq:additive}. Additional discussion regarding conceptual differences between Equations \ref{eq:additive} and \ref{eq:pan2001} are given below in Section 5.2, in the context of semi-competing events.

\textit{Estimation:} Under the gamma frailty model, Assumptions A.1--A.2 and the assumption that the censoring time, the covariates, and the random error $U_{ij}$ are mutually independent, \cite{pan2001using} derived a system of estimation equations for $\beta$, based on the linear rank statistics used in the univariate case \citep{tsiatis1990estimating} and the EM algorithm. Via a profile likelihood argument and an EM-like algorithm, estimators of $\theta$ and $\lambda_0$ are derived. The variance estimators are derived by the jackknife approach. More stable estimation procedures were proposed by \cite{zhang2007alternative,xu2010like,johnson2012smoothing}. The latter paper also considered the inverse Gaussian frailty distribution with the weighted bootstrap approach \citep{ma2005robust} for the variance estimation. However, none of the above estimators are semiparametric efficient because the EM-like algorithms do not maximize the likelihood function. Moreover, the asymptotic properties of these estimators have not been studied. 

\cite{liu2013kernel} developed a nonparametric maximum likelihood estimation method for the model of \cite{pan2001using} using the clever kernel smoothed profile likelihood approach of \cite{zeng2007efficient}. Under Assumptions A.1--A.2 and the assumption that $T^o_{ij}$ and $C_{ij}$  are independent given $Z_{ij}$, they showed that their proposed estimator for $\beta$ is consistent, asymptotically normal, and semiparametric efficient when the kernel bandwidth is properly chosen. An EM-aided numerical differentiation method is derived for variance estimation. A similar approach was adopted by \cite{chen2013estimation} with the generalized gamma distribution. 

\textit{Software:} The function \texttt{frailtyGAFT} of the \texttt{R} package \texttt{spBayesSurv} \citep{spbayes2020} fits a generalized accelerated failure time frailty model for clustered data (among other types of data) with normal frailties, and uses a Bayesian approach. The function \texttt{mlmfit} of \texttt{frailtyHL} \citep{doha2019} fits AFT models for clustered data based on the h-likelihood approach.

\section{COX-TYPE MODELS FOR CLUSTERED COMPETING RISKS DATA}
In the standard competing risks setting with no clustered data and a model based on the cause-specific hazards (Cox-type, AFT-type, or any other form), the likelihood (or partial likelihood) of $n$ independent observations factors into separate components for each failure type. Each factor is precisely the term that would be obtained by regarding other failure types as censored at the observed time \cite[Chapter~8]{kalbfleisch2011statistical}. Therefore, 
as long as there are no common parameters among the $L$ cause-specific hazards, standard maximum likelihood (or partial likelihood) can be used for estimation, regarding each event type in turn as the main event of interest and the other events as censoring. 
However, \cite{gorfine2011frailty} showed that this factorization of the likelihood no longer holds under the general setting of clustered data with competing events.

With clustered competing risks data, 
the observed data consists of $\{T_{ik},\delta_{ik},J_{ik},Z_{ik} \, ; \, i=1,\ldots,n \, , \, k=1,\ldots,m_i \}$.
\cite{gorfine2011frailty} proposed a class of frailty models with a flexibile correlation
structure among failure types within a cluster. They showed that this class of models
includes the model of \cite{bandeen2002modelling} as a special case. Specifically, for $j=1,\ldots,L$, $i=1,\ldots,n$, and $k=1,\ldots,m_i$, the cause-specific hazard functions are given by
\begin{eqnarray*}
\lambda_{jik}(t|Z_{ik},\mathcal{F}_{it}) &=& \lim_{\Delta \rightarrow 0} \Delta^{-1} \Pr(t \leq T^o_{ik} < t+\Delta, J_{ik}=j|T^o_{ik} \geq t, Z_{ik},\mathcal{F}_{it}) \\
&=& \lambda_{0j}(t) \exp\{\beta_j^T Z_{ik}+ \epsilon_{ij}(t)\} 
\end{eqnarray*}
where $\beta_1,\ldots,\beta_L$ and $\lambda_{01},\ldots,\lambda_{0L}$ are cause-specific vectors of regression coefficients and baseline hazards functions, respectively. Clearly, if it is desired not to include a given covariate in the model for a particular event type,
the corresponding coefficient can be set to 0. The association between outcomes of members of cluster $i$ is induced by the latent cluster-specific frailty process history
up to the cluster’s maximum observed time. Conditional on the cluster's frailty process history and the observed covariates, the survival times within cluster are assumed to be independent. For any given set of times $(t_{i1},\ldots,t_{im_i})$, $\{\epsilon_i(t_{i1})^T,\ldots, \epsilon_i(t_{im})^T\}$ are assumed to be independent across $i$ with density $f\{\cdot|\theta(t_{i1},\ldots,t_{im_i})\}$ for a specified parametric function $\theta(\cdot)$.

The distribution of $\{\epsilon_i(t_{i1})^T,\ldots, \epsilon_i(t_{im})^T\}$ defines the dependence between failures times of two members of the same cluster, $T^o_{ik}$ and $T^o_{ik'}$, $k \neq k'$, while within-subject dependence between failure times of different types is left unspecified, as usual in competing risks. Gorfine and Hsu provide examples for choices of $f\{\cdot|\theta(t_{i1},\ldots,t_{im_i})\}$. A practically important example is the multivariate normal distribution with time-independent frailty, i.e., $\epsilon_i(t) \equiv  (\epsilon_{i1},\ldots,\epsilon_{iL})$ is zero-mean multivariate normally distributed with an unknown covariance matrix. The covariance between $\epsilon_{ij}$ and $\epsilon_{ij'}$ reflects the dependence between times to event of types $j$ and $j'$ among any two members of cluster $i$.   

\textit{Estimation:} Parametric and semiparametric maximum likelihood estimation procedures, based on the EM algorithm,
were given in \cite{gorfine2011frailty}. Efforts have been made to simplify the estimation procedure of \cite{gorfine2011frailty} under certain frailty distributions. \cite{dharmarajan2018evaluating} considered the case of $L=2$ competing events and the time-independent multivariate normal frailty distribution. \cite{eriksson2015additive} and \cite{rueten2019investigating} used time-independent additive-gamma frailty models. \cite{do2017statistical} considered time-independent frailty and developed h-likelihood estimators. An extension of the h-likelihood approach to allow for penalized variable selection was provided by \cite{rakhmawati2021penalized}.

\textit{Software:} The \texttt{R} package \texttt{nvm\_suv}  applies the semiparametric approach of \cite{gorfine2011frailty} under the multivariate normal frailty (can be fount at \texttt{https://github.com/AsafBanana/nvm\_suv}). The package \texttt{frailtyHL} \citep{doha2019} uses the h-likelihood approach.

\section{SEMI-COMPETING RISKS VIA ILLNESS-DEATH MODELS}

A semi-competing risks model includes three stochastic processes as depicted in Figure \ref{Fig:multis}: the time to the nonterminal event, time to the terminal event without experiencing the nonterminal event, and time to terminal event after the nonterminal event. In contrast to the standard competing risks setting, semi-competing risks data includes at least partial information on the joint distribution of nonterminal and terminal events.  Consider, for example, an illness-death model with age at diagnosis of a certain disease and age at death as the respective nonterminal and terminal events. In such a setting, two issues need to be dealt with: (i) age at death after disease diagnosis is left truncated by the age at diagnosis; (ii) in most applications, it is unrealistic to assume conditional independence between age at diagnosis and age at death after diagnosis, given the observed covariates. \cite{xu2010statistical} suggested the frailty approach for dealing with (ii). We focus
on this approach here. For other illness-death models (e.g. copula, causality) see \cite{gorfine2020marginalized}, \cite{lee2021fitting}, and \cite{nevo2021}. 

Additional notation is required. Let $V_{i}=\min(T^o_{i1},T^o_{i2},C_i)$ be the time to earliest event or censoring, $\delta_{i1}=I(V_{i1}=T^o_{i1})$ and $\delta_{i2}=I(V_{i1}=T^o_{i2})$ the respective indicators of the nonterminal and terminal events, and $W_{i}=\delta_{i1} \min(T^o_{i2},C_i)$ be the time to terminal event or censoring if the nonterminal event observed 
and 0 otherwise. Finally, we define the terminal event indicator  $\delta_{i3}=\delta_{i1}I(W_i=T^o_{i2})$ for those with observed nonterminal event. The observed data consists of $\{V_{i},W_{i},\delta_{i1},\delta_{i2},\delta_{i3},Z_{i} \, ; \, i=1,\ldots,n\}$.

\subsection{Cox Models}

\subsubsection{Conditional Approach}
\cite{xu2010statistical} suggested  an illness-death model with three Cox-based hazard functions. One of their major contributions was the inclusion of a gamma-distributed shared frailty variate, that acts multiplicatively on each of the hazard functions, for incorporating unobserved dependence between the times to nonterminal and terminal events. Specifically, the hazard functions of the three transition processes are given by 
$$
\lambda_{1}(t|Z_{i},\epsilon_i) = \lim_{\Delta \rightarrow 0} \Delta^{-1} \Pr (t \leq T^o_{i1} < t+\Delta|T^o_{i1} \geq t, T^o_{i2} \geq t, Z_{i},\omega_i) =
\omega_i \lambda_{01}(t) \exp(\beta^T_1 Z_i)
$$
$$
\lambda_{2}(t|Z_{i},\epsilon_i) = \lim_{\Delta \rightarrow 0} \Delta^{-1} \Pr (t \leq T^o_{i2} < t+\Delta|T^o_{i1} \geq t, T^o_{i2} \geq t, Z_{i},\omega_i) =
\omega_i \lambda_{02}(t) \exp(\beta^T_2 Z_i)
$$
$$
\lambda_{3}(s|t,Z_{i},\epsilon_i) = \lim_{\Delta \rightarrow 0} \Delta^{-1} \Pr (s \leq T^o_{i2} < s+\Delta|T^o_{i1} = t, T^o_{i2} \geq s, Z_{i},\omega_i) =
\omega_i \lambda_{03}(s) \exp(\beta^T_3 Z_i) 
$$
where $s > t >0$, $\lambda_{0l}$ is a transition-specific baseline hazard function,
and $\beta_l$, $l=1,2,3$, are transition-specific vectors of regression coefficients. The functions $\lambda_1$ and $\lambda_2$ are the usual cause-specific hazard functions for competing events, while $\lambda_{3}(s|t,Z_{i},\epsilon_i)$ is the hazard for occurrence of the terminal event following the occurrence at time $t$ of the nonterminal event. We avoid using the sojourn time $s-t$ in the third process from nonterminal to terminal event,  since a negative association is often expected between time to the nonterminal event and the time elapsed between the nonterminal and the terminal events. For example, the residual life after
disease diagnosis tends to be shorter for those diagnosed at older ages than for those diagnosed at younger ages.

The time $T^o_{i1}=t$ to nonterminal event is not included in the vector of covariates of $\lambda_3$. Instead, the dependence between the two potential event times $T_{i1}$ and $T_{i2}$ arises from two sources: the so-called explanatory hazard ratio \citep{xu2010statistical} $\lambda_1/\lambda_3$ and the frailty variate $\omega_i$. If $T_{i1}$ and $T_{i2}$ are independent, the explanatory hazard ratio should be 1 at all time points, and the frailty variate should also be 1. 

\textit{Estimation:} \cite{xu2010statistical} showed that under gamma-distributed frailty with mean 1 and variance $\theta$, the unconditional likelihood function, after integrating out the frailty variate, has a closed form. They then provided semiparametric maximum likelihood estimators, under Assumptions A.1 and A.3 and additional regularity assumptions.
\cite{lee2015bayesian} adopted the model of  \cite{xu2010statistical} but replaced their semiparametric maximum likelihood estimation procedure with a semiparametric Bayesian estimation approach. \cite{jiang2017semi} developed a class of transformation models that permits a nonparametric specification of the frailty distribution, at the price of using parametric forms for the transformation and function and error distribution, to insure identifiabilty. They derived the semiparametric efficient score under the complete data setting and proposed a non-parametric score imputation method to handle right censoring. The proposed estimators were shown to be consistent and asymptotic normal. \cite{lee2021fitting} considered Weibull or B-spline baseline hazards with a fixed number of knots, and provided maximum likelihood estimators.

\subsubsection{Marginalized Approach}
Recently, \cite{gorfine2020marginalized}\ proposed an alternative frailty-based illness-death model with a Cox-type 
form used for the marginal hazards rather than the conditional hazard. Their work also accommodates delayed entry.
The conditional hazards are constructed in such a way that the marginal hazards are of Cox-type form.
Specifically, the conditional hazard functions are given by 
$$
\lambda_{1}(t|Z_{i},\epsilon_i) = 
\omega_i \alpha_{1}(t| Z_i) \hspace{0.3cm} t>0
$$
$$
\lambda_{2}(t|Z_{i},\epsilon_i) = 
\omega_i \alpha_{2}(t| Z_i) \hspace{0.3cm} t>0
$$
$$
\lambda_{3}(s|t,Z_{i},\epsilon_i) = 
\omega_i \alpha_{3}(s| Z_i) \hspace{0.3cm} s>t>0
$$
and the corresponding marginalized hazard functions are defined to be
$$
\lambda_{1}(t|Z_{i}) = \lambda_{01}(t) \exp(\beta^T_1 Z_i) \hspace{0.3cm} t>0
$$
$$
\lambda_{2}(t|Z_{i}) = \lambda_{02}(t) \exp(\beta^T_2 Z_i) \hspace{0.3cm} t>0
$$
$$
\lambda_{3}(s|t,Z_{i}) = \lambda_{03}(s) \exp(\beta^T_3 Z_i) \hspace{0.3cm} s>t>0
$$
with unspecified baseline hazard functions. They showed that under the gamma frailty model, the nonnegative functions $\alpha_l$, $l=1,2,3$, has a closed analytical form. In contrast to the above marginal hazard functions with standard interpretation,  the marginal hazards in the model of \cite{xu2010statistical} are of a complicated structure and depend on the parameters of
the frailty distribution \cite[Equations (19)--(21)]{xu2010statistical}. Gorfine et al.\ developed estimators for
their model using a pseudo-likelihood approach. Table \ref{tbl:methods2} summarizes the available frailty-based illness-death models, estimation procedures, and \texttt{R} packages.

\subsection{AFT models}
Unlike Cox-type models, AFT models for the illness-death framework are well developed. \cite{lee2017accelerated} proposed
the following scale-change model
\begin{eqnarray*}
	\log(T^o_{i1})&=&\beta^T_{1} Z_i+\epsilon_i+U_{i1} \, ,\, T^o_{i1}>0\\
	\log(T^o_{i2})&=&\beta^T_{2} Z_i+\epsilon_i+U_{i2} \, ,\, T^o_{i2}>0\, , \,\mbox{given subject $i$ is free of the disease}\\
	\log(T^o_{i2})&=&\beta^T_{3} Z_i+\epsilon_i+U_{i3} \, ,\, T^o_{i2}>t_{1}>0, \,\mbox{given subject $i$ was diagnosed at age $T^o_{i1}=t_1$}
\end{eqnarray*}
where the random errors $U_{il}$ are independent across $i=1,\ldots,n$ and $l=1,2,3$ with possibly unspecified distributions. Given that subject $i$ was diagnosed at age $T^o_{i1}=t_1$, the support of $T^o_{i2}$ is restricted by $t_1$, so the conditional distribution of $T^o_{i2}$ is truncated by $t_1$. The dependence between $T^o_{i1}$ and $T^o_{i2}$ is defined by the distribution of $\epsilon_i$, and takes the form of an additive term
in the log of the failure times. For the case where $\epsilon_i$
is normally distributed, \cite{lee2017accelerated} presented parametric and semiparametric estimation methods based on a Bayesian approach.  

Recently, \cite{katz2022}  provided a gamma-frailty illness-death AFT model where the frailty acts multiplicatively on the hazards of the error terms, in the spirit of \cite{pan2001using}:
\begin{eqnarray*}
	\log(T^o_{i1})&=&\beta_{1}^{T} X_{i}+U_{i1} \, ,  \, T^o_{i1}>0\\
	\log(T^o_{i2})&=&\beta_{2}^{T} X_{i}+U_{i2} \, ,  \, T^o_{i2}>0 \, ,\,\mbox{given subject $i$ is free of the disease}\\
	\log(T^o_{i2})&=&\beta_{3}^{T} X_{i}+U_{i3} \, ,  \, T^o_{i2}>t_{i1}>0 \, ,\,\mbox{given subject $i$ was diagnosed at age $T^o_{i1}=t_1$}
\end{eqnarray*}
where $U_{il}$ are random errors with unspecified distributions. The dependence between $T^o_{i1}$ and $T^o_{i2}$ is incorporated via a shared frailty model, but not directly as a linear component of the log failure times. Rather, given individual $i$'s frailty variate $\omega_i$, it is assumed that the respective conditional baseline hazard functions of $A_{il} = \exp(U_{il})$, $l=1,2,3$, are given by
\begin{eqnarray*}
	\lambda_{1}(t|\omega_i) &=& \omega_i \lambda_{01}(t) \, , \, t>0    \\
	\lambda_{2}(t|\omega_i) &=& \omega_i \lambda_{02}(t) \, , \, t>0 \, ,  \mbox{given subject $i$ is free of the disease} \\
	\lambda_{3}(s|t_1,\omega_i) &=& \omega_i \lambda_{03}(s) \, , \, s>t_1>0 \, , \mbox{given subject $i$ was diagnosed at age $T^o_{i1}=t_1$}
\end{eqnarray*}
where each $\lambda_{0j}(\cdot)$ is an unspecified baseline hazard function of $A_{jk}$ and $\omega_i$ is independent of $X_i$. It is assumed that the $\omega_i$'s are gamma distributed with mean 1 and unknown variance $\theta$. 
\cite{katz2022} extended the estimation approaches of \cite{zeng2007efficient} and \cite{liu2013kernel}, and developed a semi-parametric maximum likelihood estimators based on a kernel-smoothed likelihood combined with an EM algorithm. 

Section 2.2 of \cite{katz2022}  discusses the conceptual differences between \cite{lee2017accelerated} and \cite{katz2022}, beyond the differences between the Bayesian and frequentist approaches. Briefly, assume that the error terms follow,
for example, a normal distribution error term with mean $\mu$ and variance  $\sigma$.  It is shown that the conditional hazard functions for the healthy $\rightarrow$ disease transitions under the multiplicative  \citep{katz2022} and additive \citep{lee2017accelerated} models are given, respectively, by
$$
	\lambda_{1}(t|Z_i, \omega_i)
	= \frac{\omega_i e^{-\beta^T_{1} Z_i} }{\sigma}
	{\phi\left(  \frac{t e^{-\beta^T_{1} Z_i} -\mu}{\sigma}   \right)}
	\left\{1-\Phi\left(  \frac{t e^{-\beta^T_{1} Z_i} -\mu  }{\sigma}  \right)\right\}^{-1}
$$
and
$$
	\lambda_{1}(t|Z_i,\omega_i)=\frac{1}{\sigma t}\phi\left(\frac{\log t-\mu-\beta^T_1 Z_i-\omega_i}{\sigma}\right)
	\left\{1-\Phi\left(\frac{\log t-\mu-\beta^T_1 Z_i+\omega_i}{\sigma}\right)\right\}^{-1}
$$
where $\phi(\cdot)$ is the standard normal density and $\Phi(\cdot)$ is the standard normal CDF. 
Under the additive model, the conditional hazard could be a non-monotone function of $\omega_i$.
By contrast, under the multiplicative frailty model the conditional hazard is a monotone increasing function of $\omega_i$ 
for any error distribution. Hence, the multiplicative-frailty model can be viewed as a model
with a simpler interpretation for the contribution of the unobserved frailty effect. 


\section{RECURRENT EVENTS}

Many studies involve recurrent event data where individuals can experience an event of interest repeatedly over time.
Examples include repeated epileptic seizures, repeated infections, and repeated hospitalizations. In this context,
the individual is the cluster. Various types of censoring can occur. For example, observation on an individual
can be limited to a specified time period, or observation can be stopped after the occurrence of a specified
number of events. Let us denote the $k$-th observed event time for individual $i$ by $T_{ik}$, $k = 1, \ldots, m_i$.
If an individual is censored before experiencing any events, $m_i$ is equal to zero.
Dependence among the event times arises from unobserved individual-specific factors.

There are two main approaches to analyzing recurrent events data. The first approach involves modeling the counting
process $N_i(t) = \sum_{k=1}^{m_i} I(T_{ik} \leq t)$. The second approach involves modeling the gap times
$D_{ik} = T_{ik} - T_{i,k-1}$ (with $T_{i0}=0$). 

There are two key quantities for model specification in the counting process approach, the \textit{stochastic 
intensity function} and the \textit{rate function}. The stochastic intensity function, which is the analogue
of the hazard function for single event data, is defined in terms of a conditional expectation as
$\lambda_i(t|\mathcal{H}_{it}) = \lim_{\Delta \downarrow 0}  \Delta^{-1} E\{N_i(t+\Delta)-N_i(t)|\mathcal{H}_{it}\}$, where
$\mathcal{H}_{it}$ denotes the
observation history on individual $i$ up to time $t$. The rate function is defined in terms of an unconditional
expectation as $\mu_i(t) = \lim_{\Delta \downarrow 0} \Delta^{-1} E\{N_i(t+\Delta)-N_i(t)\}$. Clearly, 
$E\{\lambda_i(t|\mathcal{H}_{it})\}=\mu_i(t)$.
If the stochastic intensity does not depend on any random quantities from the past history,
then $N_i(t)$ is nonhomogeneous Poisson process; see Andersen et al.\ (1993, Section II.5.2).

Andersen and Gill (1982) presented the
counting process version of the \cite{cox1972regression} model, which takes 
$\lambda_i(t|\mathcal{H}_{it}) = \lambda_0(t) Y_i(t) e^{\beta^T Z_i}$, with $Y_i(t)$ being
an 0--1 variable indicating whether individual $i$ is (1) or is not (0) under observation at time $t$.
Here, we will assume that the entire trajectory of $Y_i(t)$ over time is part of the information
available at time zero. Introducing a frailty into this model gives $\lambda_i(t|\mathcal{H}_{it},\omega_i) = 
\omega_i \lambda_0(t) Y_i(t) e^{\beta^T Z_i}$. For the setting in this paper where the covariates are time-independent, 
this yields the doubly stochastic Poisson process, also known as the Cox process \citep{Cox1955}. Likelihood-based
inference for this model is discussed in \cite[Section 3.5]{cook2007}; see also  Andersen et al.\ (1993, Chapter IX)
for a discussion focusing on the gamma frailty case. Other estimation methods for the Cox frailty model
can be extended to this counting process frailty model.

In a very recent paper, \cite{bedair2021} extended the foregoing model to the case where
there are multiple types of recurrent events. The model is
$$
\lambda_{ij}(t|\mathcal{H}_{it},\omega_i) = 
\omega_{ij} \lambda_{0j}(t) Y_i(t) e^{\beta_j^T Z_i}
$$
where $j$ indexes the failure type and
$\omega_i$ is now a vector. The distribution of $\omega_i$ is specified by specifying the marginal
distribution of each $\omega_{ij}$ (as, for example, gamma or lognormal) and expressing
the dependence between the $\omega_{ij}$'s across $j$ using a copula function (Clayton-type
or multivariate normal).

\cite{box2006} introduced the model
$$
\lambda_{ik}(t|Z_i,T_{i,k-1}) = \omega_i \lambda_{0k}(t-T_{i,k-1}) e^{\beta^T Z_i}
$$
where $\lambda_{ik}(t|Z_i,T_{i,k-1})$ is hazard for experiencing the $k$-th event given the covariates and the time
of the $(k-1)$-th event and $\lambda_{0k}$ is a baseline hazard for the $k$-th event. Note that the hazard model
here is expressed in term of $t-T_{i,k-1}$, the time elapsed from the $(k-1)$-th event. The frailty $\omega_i$
was taken to be gamma-distributed. Estimation was based on penalized likelihood, similar to \cite{therneau2003penalized}.

A number of papers have considered joint frailty models for recurrent events in conjunction with a terminal event.
In some cases, the time to the terminal event is of interest in its own right, whereas in other cases it is regarded 
as an informative censoring time. One example of the first case is the work of \cite{Maz2012}, who present the model
\begin{align*}
\lambda_{Ri}(t|\mathcal{H}_{it}, \omega_i, \eta_i) & = \omega_i \eta_i \lambda_{R0}(t) Y_i(t) e^{\beta_R^T Z_i} \\ 	
\lambda_{Ti}(t|\mathcal{H}_{it},\omega_i) & = \omega_i  \lambda_{T0}(t) Y_i(t) e^{\beta_T^T Z_i}  	
\end{align*}
where $\omega_i$ and $\eta_i$ are frailty variates. 
Here, $\lambda_{Ri}$ denotes the stochastic intensity of the recurrent event process and $\lambda_{Ti}$ denotes the
hazard of the terminal event.
Mazroui et al.\ assumed that $\omega_i$ and $\eta_i$
are independent gamma random variables with different shape parameters.
\cite{liulei2016} considered a joint frailty model for recurrent events and a terminal event where the
recurrent event is subject to zero-inflation.

An example of a case where the time to the terminal event is regarded as informative censoring time
is the work of \cite{wang2001}.
They assume that the recurrent event process is a nonhomogeneous Poisson process with frailty, and
conditional on the frailty, the terminal event is independent of the recurrent event process.
They present an estimator based on an estimating equation involving inverse probability weighting.

Regarding models based on gap times, a basic model is to assume that the conditional on the frailty, each individual's
gap times are independent and identically distributed, so that they form a renewal process. Inference 
for such models can be carried out by maximum likelihood. These models are discussed in Section 4.2.2
of \cite{cook2007}.

Finally, we discuss the piecewise-exponential additive mixed effects model (PAMM) of \cite{bender2018}.
For single event data, the model is defined as follows. We divide the time axis into intervals $(\kappa_{j-1},\kappa_j]$.
Associated with each interval $j$ is a selected point $t_j$ within the interval, such as the midpoint or one of the
endpoints. Associated with each cluster $i$ is a normally-distributed random effect $\epsilon_i$. The conditional
hazard given the random effect term is then specified as
$$
\lambda_{ik}(t|\epsilon_i)
= \exp \left( f_0(t_j) + \sum_{r=1}^p f_r(Z_{ikr},t_j) + \epsilon_i  \right), \quad t \in (\kappa_{j-1},\kappa_j]
$$
where $f_0$ is modeled as a spline and $f_r$ is modeled as a two-dimensional product tensor spline.
The piecewise-exponential structure allows the likelihood of the model to be expressed in the form
of the likelihood in a generalized additive mixed model (GAMM) with a Poisson response.
Estimation is carried out using the approach of \cite{wood2011}, which is based on penalized likelihood where the spline 
coefficients are regarded as multivariate normal and a Laplace approximation is used to evaluate 
the integrals in the likelihood function. Software for carrying out the estimation is available
in the \texttt{R} package \texttt{pammtools} \citep{pamm}. The estimation procedure can be applied to
recurrent events data by preprocessing the dataset in a suitable way. Details of the preprocessing
for the model based on gap times are described on the following webpage:

\url{https://adibender.github.io/pammtools/articles/recurrent-events.html}

\textit{Software}: Several \texttt{R} functions can handle recurrent events data. The \texttt{coxph} function in the \texttt{survival}
package \citep{survival-package} and the \texttt{coxme} function in the \texttt{coxme} package \citep{therneau2020coxme} 
can handle Cox-type models for recurrent event data if
counting process input is used. The \texttt{reda} package \citep{reda-package} handles recurrent events data with gamma
frailty. In addition, as described in the text, the \texttt{pammtools} package \citep{pamm} can be used to implement the method of \cite{bender2018}.

\section{RANDOM SURVIVAL FORESTS FOR CLUSTERED DATA}

Methods based on recursive partitions, such as regression trees \citep{breiman1984classification}, provide
useful nonparametric alternatives to parametric or semiparametric regression methods. Starting with the full dataset at the root node, each step of a recursive partitioning algorithm involves splitting a subset of the original dataset into two descendant subsets, known as daughter nodes. The dataset is split based on a covariate $Z$ and a threshold value $c$, so that the left daughter node contains observations in the current data subset with $Z < c$ and the right daughter contains observations in the current data subset with $Z \geq c$. The splitting covariate and threshold value are chosen to maximize the difference between the two descendant datasets with respect to an outcome variable, based on a pre-specified criterion.
In our context, a good split for a node maximizes the survival difference between daughters.
The splitting covariate can be either a single covariate from the original dataset 
or a linear combination of covariates. The process is repeated
recursively for each subsequent node. Descendant datasets that do not include
a prespecified minimum number observations or events are not split any further; these datasets are referred to as
terminal nodes or leaves of the tree. As the number of nodes increases, each terminal node is expected to be homogeneous and is populated by observations with similar survival times. 

Predictions for new observations are computed by determining which terminal node each new observation falls into and then aggregating the outcomes of participants in the training data who were mapped to the same node. Another important feature of regression trees is a fully nonparametric measure of variable importance (VIMP).

It is well known that constructing ensembles composed of several base learners, such as trees, can substantially
improve prediction performance. Moreover, \cite{breiman2001random} showed that ensemble learning can be improved further by injecting
randomization into the base learning process, an approach called \textit{random forests}. Random forest combines a large
number of trees by taking the same number of bootstrap samples from the original data, for each bootstrap tree. In addition,
at each split, the search of the best splitting covariate is conducted only from a random subset of covariates. From all trees created in this process, a ``forest" is created. The response variable value of a new observation
based on a random forest of $B$ trees is predicted by dropping the vector of covariates down each tree, and averaging over the $B$ terminal nodes. 

\cite{gordon1985tree} were the first to adapt the Classification and Regression Trees (CART) method \cite{breiman1984classification} to censored data. 
Since then, many methods for regression survival trees, random survival forests, and measures of VIMP, for uncorrelated survival data without or with competing events, have been developed; a partial list includes \cite{hothorn2006unbiased,ishwaran2008random,zhu2012recursively,ishwaran2014random,ishwaran2019standard}, and a recent review is provided by \cite{wang2019machine}. 
Here we provide a review of survival trees and random survival forests for clustered data.  

A regression tree algorithm consists of three building blocks: (i) a splitting rule; (ii) a pruning rule that determines the proper tree size; and (iii) a statistic that summarizes the properties of the terminal nodes. For clustered data, 
\cite{su2001multivariate},  \cite{gao2006developing} and \cite{fan2006trees}  proposed algorithms based on
marginal Cox proportional hazard model. The splitting rule is based on
the robust logrank statistic obtained using the marginal modeling approach of \cite{wei1989regression}. The rest of the tree procedure proceeds as standard univariate survival trees, with the summary at the terminal nodes again based on marginal analysis (e.g., separate Kaplan-Meier curves for each terminal node).
 
\cite{gao2004identification} proposed a method that extends the CART algorithm to clustered survival data
by introducing a gamma frailty to account for dependence among correlated events.  For each nonterminal node $\psi$, and for each potential splitting covariate and threshold value $c$, Gao et al.\ use the extended Cox proportional hazards model
$$
\omega_i \lambda_{0\psi,q}(t) \exp\{\beta_{\psi,q} I(Z_{qik} \geq c)\}   \hspace{0.4cm} ik \in \psi  \hspace{0.4cm} q =1, \ldots, p 
$$
where $\omega_i$ has a gamma distribution with mean 1 and unknown node-dependent variance $\theta_{\psi}$ and 
the hazard functions $\lambda_{0\psi,q}(t)$ are of unspecified form. For each potential splitting covariate $Z$ and threshold value $c$, $(\beta_{\psi},\theta_{\psi})$ is estimated based on the penalized partial likelihood approach of \cite{therneau2003penalized}. Defining $\widehat{\beta}_{\psi}$ to be the resulting estimator and  $\widehat{V}(\widehat{\beta}_{\psi})$ to be its estimated variance, the splitting criterion is the Wald-test statistic
$$
\max_{q} \max_{c} \widehat{\beta}^{\, 2}_{\psi,q} /\widehat{V}(\widehat{\beta}_{\psi,q}) \, .
$$
For pruning the tree, the authors used the bottom-up algorithm; see Section 4.2 of \cite{segal1988regression}. Finally, Kaplan-Meier survival curves are computed for each terminal node and  used for survival prediction of new observations. At about the same time, \cite{su2004multivariate}  also used a gamma-frailty survival tree method, with a splitting rule based on the maximized integrated log likelihood, and a pruning rule based on the Akaike information criterion (AIC). 

The above frailty-based methods suffer from two main drawbacks. First, the computation time is extremely long with large data sets. Second, in settings where all cluster members share the same censoring times, the estimation procedure often fails to converge. See \cite{levine2014bayesian} and \cite{hallett2014random} for further discussion. \cite{fan2009multivariate} suggested to overcome these difficulties by replacing the unspecified hazard functions with $\exp(\beta_{0\psi})$. Alternatively, \cite{levine2014bayesian} used a gamma frailty model with a piecewise constant baseline hazards, and developed Bayesian multivariate survival tree methods that eliminate the foregoing two obstacles. \cite{hallett2014random} used the model of \cite{fan2009multivariate} for building a random forest and studied various variable importance measures.


\section{CONCLUDING REMARKS}

In this review we summarized the available advanced frailty-based survival models, methods and software, which also helped us revealing gaps for future projects. This includes:
\begin{itemize}
	\item Semiparametric AFT models for clustered data and a single outcome with various frailty distributions, other than gamma. Clearly,  it would require the use of numerical integration.
	\item Frailty-based AFT models for clustered data with competing events. For example, one can start with the likelihood of \cite{gorfine2011frailty}, use a multivariate frailty setting and combine it with the smoothed kernel approach of \cite{zeng2007efficient} and \cite{liu2013kernel}.
	\item Semicompeting risks via illness-death AFT models with other frailty distributions beside gamma. 
\end{itemize}
We also believe that there could be a place for additional investigation in the area of random survival forests for clustered data, since existing works that uses frailty models focus on extended Cox models with gamma frailty distribution. Moreover, \texttt{R} packages that apply RSF for clustered survival data are currently missing.

Another important aspect is model diagnostic. It would be ideal if one could choose the frailty distribution as close to the true distribution
as possible to ensure a valid inference. But in random effects models it could be a challenging task due to lack of sufficient data to discern various models. Model diagnostic procedures have been developed; a partial list is given at the end of Section 3.1. Also, \cite{li2021model}  recently proposed to diagnose censored regression with randomized survival probabilities, a general tool that can be easily modified to various survival setting (e.g. clustered data). \cite{katz2022} applied their approach in the setting of illness-death models.

\section*{DISCLOSURE STATEMENT}
The authors are not aware of any affiliations, memberships, funding, or financial holdings that
might be perceived as affecting the objectivity of this review. 

\section*{ACKNOWLEDGMENTS}
The work of MG was supported by the Israel Science Foundation (ISF) grant number 767/21 and a grant from the Tel Aviv University Center for AI and Data Science (TAD).

%




\newpage

\begin{center}
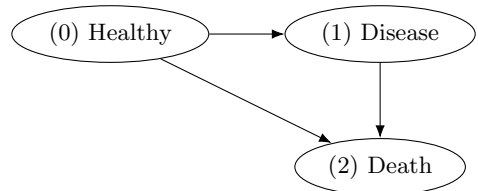

	\begin{tikzpicture}
		\node[state, ellipse] (1) at (0,0) {(0) Healthy};
		\node[state, ellipse] (2) [right = of 1] {(1) Disease};
		\node[state, ellipse] (4) [below = of 2] {(2) Death};		
		\path (2) edge (4) ;
		\path (1) edge  (2);
		\path (1) edge (4);
	\end{tikzpicture}\vspace*{0.2cm}
	\captionof{figure}{{A Semi-competing risks setting of a chronic disease.}}\label{Fig:multis}
\end{center}

\begin{table}
	\centering
	{\footnotesize
	\caption{Clustered data with a single event: \texttt{R} packages availability.}	\label{tbl:methods1}
	\begin{center}
		\begin{tabular}{p{3.3cm} p{2.1cm} p{2.2cm} p{2cm} p{3cm}}
			\hline
			 Authors & Model & Frailty Model & Procedure & package::function \\
			\hline
\cite{balan2019} & Cox, US $\lambda_0$ & G, PS, PVF &  EM algorithm & frailtyEM::emfrail \\
\cite{Belitz2017} & Cox, US $\lambda_0$ & LN & penalized PL & R2BayesX::bayesx\\
\cite{Gu2014} & Cox, US $\lambda_0$ & LN & penalized PL & gss::sscox \\
\cite{doha2019} &  Cox, S or US $\lambda_0$ & G, LN & h-likelihood & frailtyHL::frailtyHL\\
\cite{monaco2018} & Cox, US $\lambda_0$ & G, LN, IG, PVF & pseudo FL & frailtySurv::fitfrail \\
\cite{munda2012} & Cox, S $\lambda_0$&  G, PS, IG & ML & parfm::parfm \\  
\cite{rondeau2019} & Cox, S or US $\lambda_0$ & G, LN & penalized PL & frailtypack::frailtyPenal \\ 
\cite{survival-package} & Cox, US $\lambda_0$& G, LN, log $t$ &  penalized PL & survival::coxph \\
\cite{therneau2020coxme} &  Cox, US $\lambda_0$ & LN & penalized PL & coxme::coxme \\ 
			\hline
			\multicolumn{5}{l}{US=unspecified, S=specified}\\
			\multicolumn{5}{l}{G=gamma, LN=log normal, PS=positive stable, PVF= power variance family, IG = inverse Gaussian}\\
			\multicolumn{5}{l}{PL=partial likelihood, FL=full likelihood, ML=marginal likelihood, LS=least squares}\\
			\hline
\end{tabular}
\end{center}}
\end{table}

\begin{table}
	\centering
	\caption{Semi competing risks via illness death models: methods and \texttt{R} packages availability.}\label{tbl:methods2}
	\begin{center}
		\begin{tabular}{ll p{3.3cm} p{1.3cm}  p{3cm}}
			\hline
			Authors & Model & Frailty Models and More & Procedure & \texttt{R} package \\
			\hline
			\cite{xu2010statistical} & Cox& G, SP & SPMLE & - \\
			\cite{lee2015bayesian} & Cox& G, SP & Baysian & SemiCompRisks \\
			\cite{jiang2017semi} & TM & NPF, known TM, P    & SES & - \\
		\cite{lee2017accelerated} & AFT& N, P, SP & Bayesian & SemiCompRisks \\
			\cite{gorfine2020marginalized} & Cox-M& G, SP & PsL & https://github.com/\newline{}nirkeret/frailty-LTRC \\
				\cite{lee2021fitting} & Cox& G, P & MLE & - \\
			\cite{katz2022} & AFT& G, SP & SPMLE & https://github.com/\newline{}LeaKats/semicompAFT \\
			\hline
			\multicolumn{5}{l}{TM=transformation model, Cox-M= marginalized Cox, G=gamma, N=normal, SP=semiparametric model}\\
			\multicolumn{5}{l}{SPMLE=semiparametric maximum likelihood estimator}\\
			\multicolumn{5}{l}{NPF=nonparametric frailty, P=parametric survival model}\\
			\multicolumn{5}{l}{SES=semiparametric efficient score, PsL=pseudo-likelihood approach}\\
			\hline
	\end{tabular}
  \end{center}
\end{table}

\newpage



\end{document}